\documentclass[aps,prl,10pt,twocolumn,floatfix]{revtex4-1}
\usepackage{amsmath,amssymb,bm,graphicx}
\usepackage{times}

\usepackage{hyperref}
\hypersetup{
colorlinks=true,
linkcolor=blue,           
citecolor=blue,           
filecolor=magenta,        
urlcolor=cyan
}

\newcommand{\papertitle}{Dissipation-induced instabilities of a spinor Bose-Einstein condensate inside an optical cavity}
\newcommand{\authornames}{E.~I.~Rodr\'iguez Chiacchio and A.~Nunnenkamp}
\newcommand{\tcm}{Cavendish Laboratory, University of Cambridge, Cambridge CB3 0HE, United Kingdom}

\begin{document}

\title{\papertitle}
\author{E.~I.~Rodr\'iguez Chiacchio and A.~Nunnenkamp}
\affiliation{Cavendish Laboratory, University of Cambridge, Cambridge CB3 0HE, United Kingdom}

\date{\today}

\begin{abstract}
We investigate the dynamics of a spinor Bose-Einstein condensate inside an optical cavity, driven transversely by a laser with a controllable polarization angle.
We focus on a two-component Dicke model with complex light-matter couplings, in the presence of photon losses.
We calculate the steady-state phase diagram and find dynamical instabilities in the form of limit cycles, heralded by the presence of exceptional points and level attraction.
We show that the instabilities are induced by dissipative processes which generate non-reciprocal couplings between the two collective spins.
Our predictions can be readily tested in state-of-the-art experiments and open up the study of non-reciprocal many-body dynamics out of equilibrium.
\end{abstract}

\maketitle

\textit{Introduction.---} Ultracold atomic gases loaded into optical cavities form an ideal set-up for the study of quantum many-body systems far from equilibrium \cite{Review}. Their large cooperativity allows reaching the strong light-matter coupling regime \cite{Brennecke2007,Colombe2007} and cavity photon losses enable in-situ monitoring of the many-body dynamics in real time \cite{MekhovNature2007,Chen2007}.
A representative example of this idea is the experimental realization of the Dicke superradiant phase transition \cite{Dicke,HeppLieb,WangHioe}, using the motional degrees of freedom of a Bose-Einstein condensate (BEC)  \cite{Esslinger,HemmerichPNAS2014}, which provided access to the observation of critical phenomena \cite{EsslingerDickeRoton} and driven-dissipative dynamics \cite{EsslingerDickeFluct}.
This was also discussed in several theoretical works \cite{DomokosPRL2010,Bhaseen1,DomokosPRA2011,Oztop2012,Bhaseen2}.
Further advances have led to the study of competition between short- and long-ranged interactions using optical lattices \cite{HemmerichPRLdec2015, Esslinger2, EsslingerMet}, the simulation of continuous symmetry breaking with multiple cavities \cite{Esslinger3,Esslinger4}, and the observation of complex many-body phenomena in multi-mode cavities \cite{Lev,Lev2,KeelingLevPRX}.

Recently, considerable progress has been made, both theoretically \cite{Mivehvar2017,Mivehvar2019} and experimentally \cite{Zhiqiang2017, EsslingerSpin, Kroeze2018,Norcia2018,Davis2019}, on the coupling of multiple internal atomic states to the cavity modes, given its potential for quantum simulation of magnetism \cite{Mivehvar2019} and for quantum-enhanced metrology \cite{Norcia2018,Davis2019}. The focus of these studies has been however on the coherent effects of the coupling, leaving the impact of dissipative processes largely unexplored. Dissipation can have noticeable effects on the properties of many-body systems, such as modifying the nature of phase transitions \cite{Diehl2,Diehl}, the form of the phase diagram \cite{Soriente,Kirton2017}, their dynamical evolution \cite{Lang2016,Zheng2018,Chiacchio2018,Buca2018}, or giving rise to topological effects \cite{Diehl2011}. Therefore, it is exciting to explore the impact of dissipation on these complex systems.

\begin{figure}[t]
\centering
\includegraphics[width=\columnwidth]{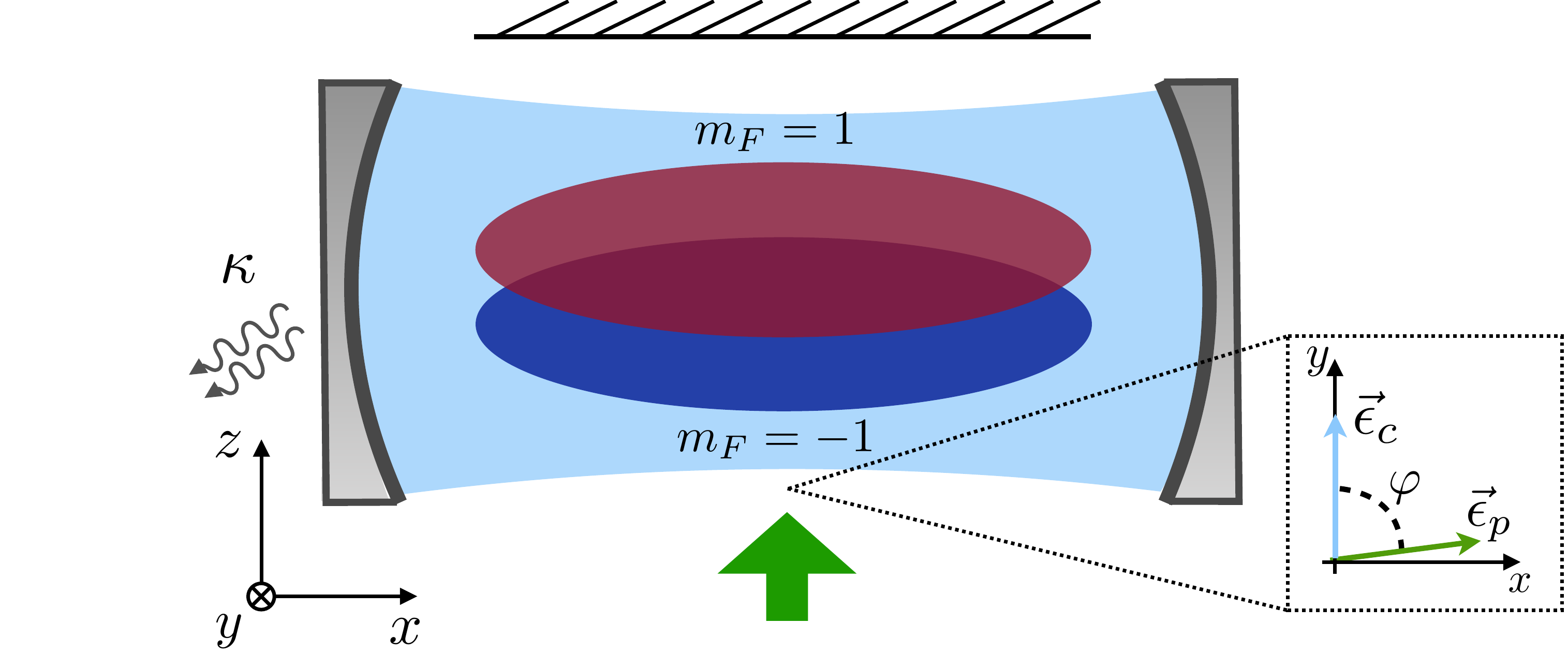}
\caption{\label{Model} Spinor Bose-Einstein condensate composed of two hyperfine states $m_{F}=\pm1$, coupled to a single-mode optical cavity with photon loss rate $\kappa$ and transversely driven by a laser whose polarization vector $\vec{\epsilon}_{p}$ is at an angle $\varphi$ with respect to the cavity field polarization vector $\vec{\epsilon}_{c}$. This leads to a finite contribution from the vectorial polarizability of the atoms, resulting in complex light-matter couplings, of equal strength but opposite phase, between the hyperfine states and the cavity field \cite{EsslingerSpin}.}
\end{figure}

In this paper we investigate the driven-dissipative dynamics of a spinor BEC composed of two hyperfine states coupled to a single mode of an optical cavity, as experimentally realized in Ref.~\cite{EsslingerSpin}, see Fig.~\ref{Model}.
This can be captured by an open two-component Dicke model with complex light-matter couplings.
We uncover the emergence of a novel unstable region that, as we show, is induced by the photon losses.
Our results are to be seen in contrast to previous studies where this type of dissipation leads to only minor quantitative changes in the phase diagram \cite{Dimer2007, DomokosPRL2010, Bhaseen1, Bhaseen2, DomokosPRA2011, Oztop2012}.
By adiabatically eliminating the cavity field, we find that the interplay between dissipation and complex coupling results in level attraction between eigenfrequencies and the appearance of anti-damping, with the emergence of instabilities being heralded by the presence of exceptional points in the spectrum.
In the unstable region, the anti-damping prevents the system from approaching a stable steady-state fixed point and leads to limit-cycle oscillations in the long-time limit. 
We trace this complex phenomena back to dissipative processes of the cavity field mediating non-reciprocal interactions between the spins.
Going beyond adiabatic elimination, we find cavity fluctuations to generate an additional anti-damping contribution that renders the normal phase unstable. Nevertheless, we show that this contribution remains negligible for typical parameters in the current generation of experiments \cite{EsslingerSpin,Kroeze2018}, allowing for observation of the aforementioned phenomena.

\textit{Model.---} We consider a gas of ultracold spin-1 atoms forming a BEC inside an optical cavity, see Fig.~\ref{Model}.
The atoms are coupled to a single cavity mode via a linearly-polarized laser that pumps the system transversely.
The atoms mediate two-photon scattering processes between the cavity and the pump which lead to transitions between the BEC state $|k_{0}\rangle$ and the excited states $|\vec{k}_{\pm,\pm}\rangle=|\pm(\vec{k}_{c}\pm\vec{k}_{p}) \rangle$, where $\vec{k}_{c,p}$ are cavity and pump momenta, respectively. We fix $|\vec{k}_{c}|=|\vec{k}_{p}|=k$ and, in this case, all the states $|\vec{k}_{\pm,\pm}\rangle$ are degenerate, thus for each atom $i$ the transitions take place between $|k_{0}\rangle^{i}$ and the symmetric state $|k\rangle^{i}=\frac{1}{2}\sum_{\mu,\nu=\pm}|\vec{k}_{\mu,\nu}\rangle^{i}$ \cite{Esslinger}.
This allows for a description of the system in terms of collective spin operators, which, in the rotating frame of the pump, reads ($\hbar=1$) \cite{EsslingerSpin}
\begin{equation}
\label{eq:er0002}
\hat{H}=-\Delta \hat{a}^{\dagger} \hat{a} + \sum_{m_{F}} \omega_0\hat{J}_{z,m_{F}}
+\frac{\hat{J}_{x,m_{F}}}{\sqrt{N_{m_{F}}}}(\lambda^{*}_{m_{F}}\hat{a} + \lambda_{m_{F}}\hat{a}^{\dagger}),
\end{equation}
where $\hat{a}$ is the bosonic annihilation operator for the cavity field, $\Delta= \omega_{p}-\omega_{c}$ is the detuning between the cavity $\omega_{c}$ and the pump $\omega_{p}$ frequency. The operator $\hat{J}_{\alpha,m_{F}}=\sum_{i} \hat{\sigma}^{i}_{\alpha,m_{F}}$ is a collective spin operator, where $\hat{\sigma}_{z,m_{F}}=|k\rangle^{i} {}^{i}\langle k|-|k_{0}\rangle^{i} {}^{i}\langle k_{0}|$ and $\hat{\sigma}^{i}_{x,m_{F}}=\frac{1}{2}(|k_{0}\rangle^{i} {}^{i}\langle k| + \textrm{H.c.})$.
The level splitting $\omega_{0}$ equals twice the recoil frequency $\omega_{r}=k^{2}/2m$, and $N_{m_{F}}$ is the number of atoms in spin state $m_{F}$.
The third term in (\ref{eq:er0002}) describes the scattering of a pump photon into the cavity mode which is accompanied by an atomic transition.
Misalignment between pump and cavity polarizations induces a non-vanishing vectorial component in the atomic polarizability, so the spin states couple differently to the cavity \cite{EsslingerSpin}.
The complex light-matter coupling $\lambda_{m_{F}}=|\lambda_{m_{F}}|e^{i\phi_{m_{F}}}$ have modulus $|\lambda_{m_{F}}|=\sqrt{\lambda_{s}^{2}\cos^2\varphi + \lambda_{v}^{2} m_{F}^2 \sin^2\varphi}$, where $\lambda_{s,v}$ is proportional to the scalar and vectorial atomic polarizabilities and $\varphi$ the angle between the pump and cavity polarization vectors $\vec{\epsilon}_{p}$ and $\vec{\epsilon}_{c}$, and $\tan \phi_{m_{F}}=\frac{\lambda_{v}m_{F}}{\lambda_{s}}\tan\varphi$ \cite{EsslingerSpin}.

For the remainder of this paper, we will focus on the case $N_{\pm 1}=N$, $|\lambda_{\pm 1}|= \lambda$ and $\phi_{1}=-\phi_{-1}=\phi$.
We obtain a two-component variant of the Dicke model \cite{Dicke, HeppLieb,WangHioe}. We stress that, for the Hamiltonian (\ref{eq:er0002}), the two effective atomic spins cannot be encapsulated in a single collective spin due to the phase difference between the couplings and that the phase difference $\phi$ cannot be removed from the Hamiltonian by any gauge transformation. Indeed, we find that it is one of the key ingredients for the effects we discuss below.

\begin{figure}[t]
\centering
\includegraphics[width=\columnwidth]{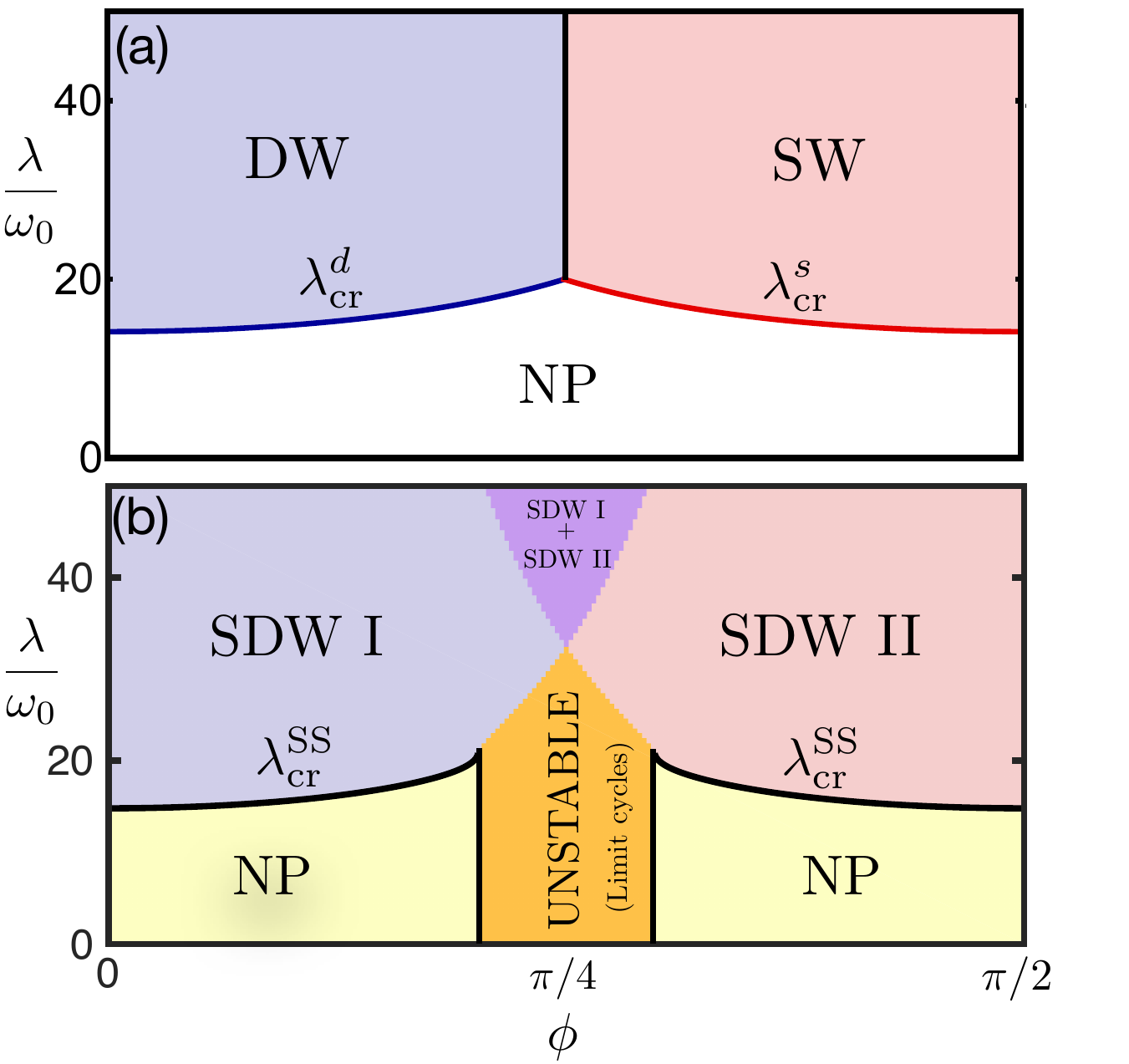}
\caption{\label{PDOrdPar}(a) Ground-state phase diagram, obtained from mean-field theory, and (b) Steady-state phase diagram, determined by the semi-classical equations of motion and a linear stability analysis \eqref{eq:er0012}, as a function of the light-matter coupling $\lambda$ and phase $\phi$, for $\Delta=-400\omega_{0}$ and $\kappa=250\omega_{0}$.
Going beyond adiabatic elimination and including the cavity field fluctuations renders the NP unstable (light orange shading) for $\phi \neq 0, \pm \frac{\pi}{2}$.}
\end{figure}

The Hamiltonian \eqref{eq:er0002} possesses a $\mathbb{Z}_{2}$ symmetry, associated with invariance under the transformation $\hat{\mathcal{U}}=e^{i\pi\hat{\mathcal{N}}}$, with $\hat{\mathcal{N}}= \hat{a}^{\dagger}\hat{a} + \sum_{\sigma=\pm1} \hat{J}_{z,\sigma}$, which can be understood as parity conservation of the total number of excitations in the system. For $\phi=0$, spontaneous breaking of this symmetry results in the well-known superradiant phase transition of the Dicke model \cite{Dicke,HeppLieb,WangHioe}, where the global spins acquire a finite and equal $x$-component. For the atom-cavity system, this corresponds to a transition from the BEC state with no photons inside the cavity, corresponding to the normal phase (NP), into a self-organized, density-wave state (DW), accompanied by the emergence of a macroscopic cavity field \cite{DomokosPRL2010, Bhaseen1, DomokosPRA2011, Oztop2012, Bhaseen2}. It has been shown in Ref.~\cite{EsslingerSpin} that for $\phi \neq 0$, the spontaneous breaking of the $\mathbb{Z}_{2}$ symmetry can lead to a different kind of superradiant order where the $x$-component of the collective spins anti-align, corresponding to each cloud of atoms self-organizing in opposite checkerboard patterns, i.e.~formation of a spin wave (SW). In Fig.~\ref{PDOrdPar}(a) we show the phase diagram obtained within mean-field theory \cite{SM}, in agreement with the observations reported in \cite{EsslingerSpin}, where the NP-DW as well as the NP-SW boundaries are given by $\lambda_{\textrm{cr}}^{d/s}=\sqrt{\frac{(-\Delta)\omega_{0}}{1\pm\cos{(2\phi)}}}$ and the DW-SW boundary is located at $\phi = \pi/4$. In the following we will study how this phase diagram is modified when taking into account the dissipative nature of the cavity.

\textit{Steady-state phase diagram.---}
We start by including dissipation in our model via a Lindblad master equation of the form $\partial_{t}\hat{\rho}=(-i)[\hat{H},\hat{\rho}] + \kappa(\hat{a}\hat{\rho}\hat{a}^{\dagger}-\frac{1}{2}\hat{a}^{\dagger}\hat{a}\hat{\rho} - \frac{1}{2}\hat{\rho}\hat{a}^{\dagger}\hat{a})$, where $\kappa$ is the photon loss rate.
We first focus on the bad-cavity limit, $(|\Delta|,\kappa) \gg (\omega_{0},\lambda)$, as studied experimentally in Refs.~\cite{EsslingerSpin, Kroeze2018}. In this limit, the cavity evolves much faster than the atoms, allowing us to adiabatically eliminate the cavity by considering the  cavity field amplitude $\alpha=\langle \hat{a} \rangle$ to be in the steady state $\alpha \approx  \frac{\lambda}{\sqrt{N}}\frac{S_{x,1} e^{i\phi}+S_{x,-1} e^{-i\phi}}{\Delta + i\frac{\kappa}{2}}$, with $S_{\sigma,\pm 1} = \langle \hat{J}_{\sigma,\pm 1} \rangle$.
By setting $\partial_{t} S_{\sigma, \pm 1}=0$ and factorizing higher-order correlations, we obtain a set of algebraic equations for the steady-state solutions \cite{SM}.
To construct the phase diagram, we determine the stability of these solutions by linearizing the equations of motion around the steady state $\hat{J}_{\sigma,\pm 1}(t) \simeq S_{\sigma,\pm 1} + \delta \hat{J}_{\sigma, \pm 1}(t)$
\begin{equation}
\label{eq:er0012}
\begin{pmatrix}\delta \dot{\hat{J}}_{x, 1} \\  \delta \dot{\hat{J}}_{y,1} \\ \delta \dot{\hat{J}}_{x,-1} \\  \delta \dot{\hat{J}}_{y,-1}\end{pmatrix}=
\begin{pmatrix} 0 & -\omega_{0} & 0 & 0 \\ \omega_{0}+\xi_{+}  & 0 & \chi_{+} & 0 \\ 0 & 0 & 0 & -\omega _{0} \\ \chi_{-} & 0 & \omega_{0} + \xi_{-} & 0 \end{pmatrix}
\begin{pmatrix} \delta \hat{J}_{x,1} \\  \delta \hat{J}_{y,1} \\ \delta \hat{J}_{x,-1} \\  \delta \hat{J}_{y,-1} \end{pmatrix}.
\end{equation}
We observe that the effects of the eliminated cavity field are that of introducing a frequency splitting $\xi_{\pm}$ between the spin components, and inducing effective interactions between the spins $\chi_{\pm}$, with both $\xi_{\pm}$ and $\chi_{\pm}$ functions of the external and order parameters.
Note that these are in general different $\chi_{+} \neq \chi_{-}$, resulting in a non-reciprocal coupling.
This means that the spins respond differently to the motion of the other one, which turns out to have a strong impact on the driven-dissipative dynamics of the system. The resulting phase diagram is shown in Fig.~\ref{PDOrdPar}(b), where we identify five different phases, classified by the order parameters $S_{x,\pm1}$ and the number of stable solutions, which we will now describe in turn. 

\begin{figure}[t]
\centering
\includegraphics[width=\columnwidth]{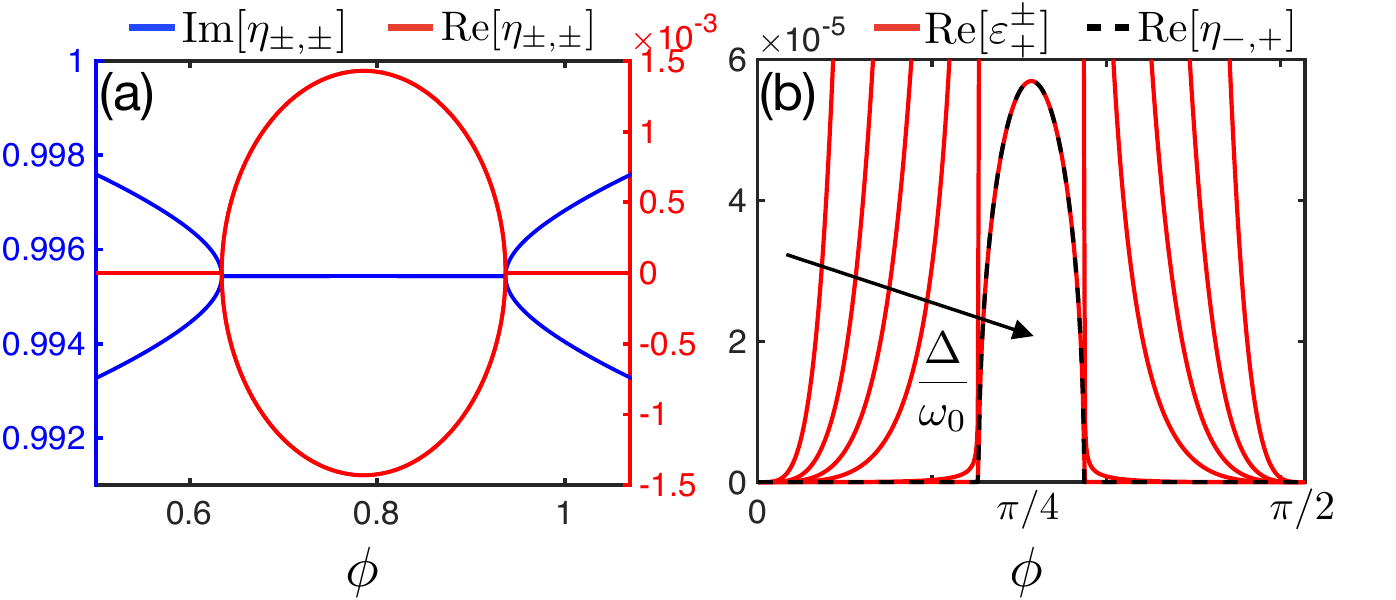}
\caption{\label{Spectrum}(a) The imaginary and real part of the eigenvalues $\eta_{\pm,\pm}$ of the dynamical matrix \eqref{eq:er0012}, resulting from adiabatic elimination of the cavity field, for $\lambda=2\omega_{0}$, $\Delta=-400\omega_{0}$ and $\kappa=250\omega_{0}$. We observe level attraction in the spectrum, consequence of the emergence of exceptional points.
(b) The real part of the pair of eigenvalues $\varepsilon^{\pm}_{+}$ (solid lines) responsible for anti-damping in the NP, obtained from the full dynamical matrix including cavity field fluctuations,
for $\Delta/\omega_{0}=10,25,50,400,1000,10000$, $\kappa=0.625|\Delta|$ and $\lambda=2\omega_{0}$. As the bad-cavity limit is approached, the eigenvalues reduce to $\eta_{-,+}$ (dashed lines) given by (\ref{eq:er0014}). All quantities are in units of $\omega_{0}$.}
\end{figure}

\begin{figure}[t]
\centering
\includegraphics[width=\columnwidth]{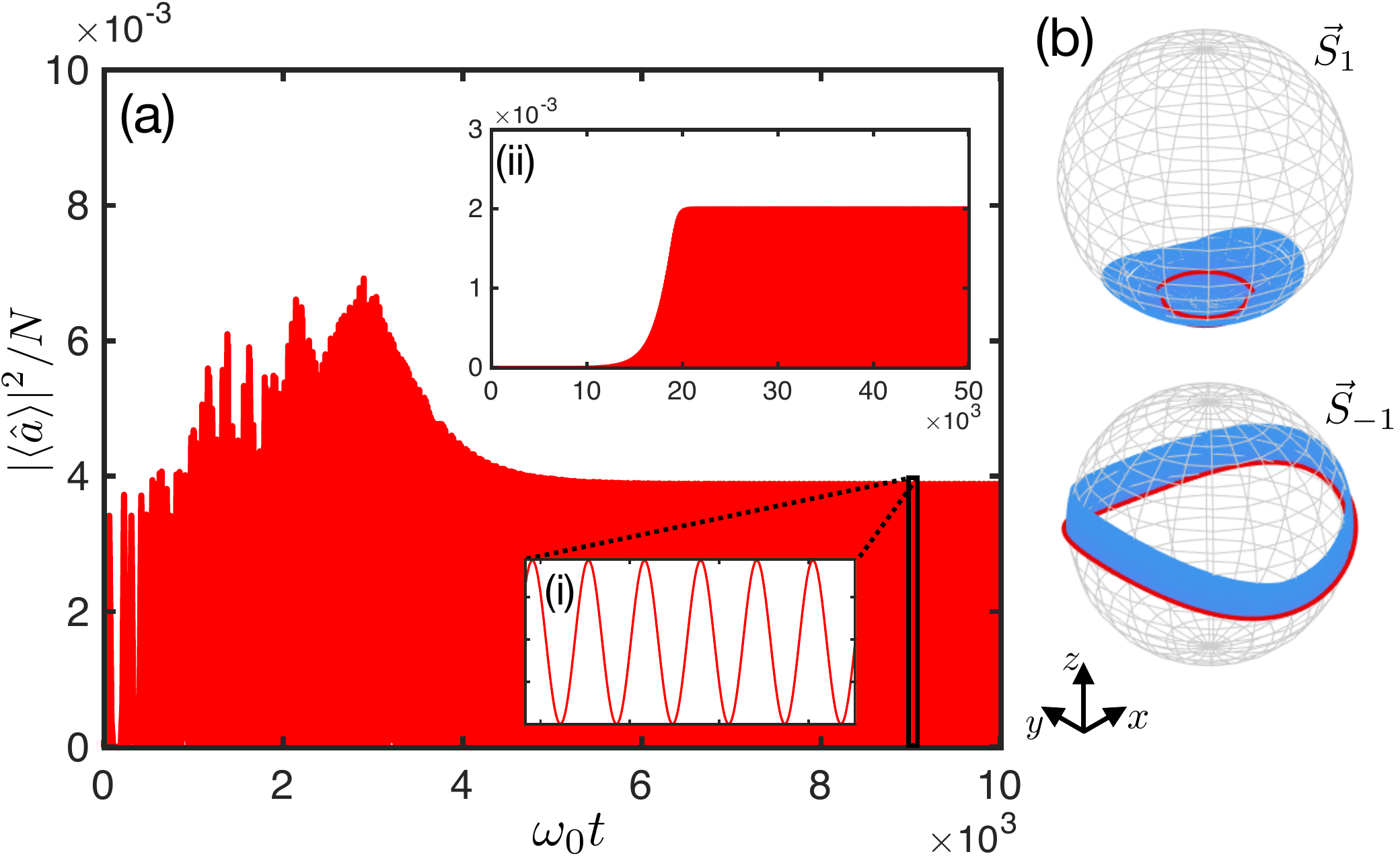}
\caption{\label{Dynamics} Dynamics in the unstable regime, with $\Delta=-40\omega_{0}$, $\kappa=25\omega_{0}$, $\lambda=5\omega_{0}$ and $\phi=\pi/4$. (a) Time evolution of the average photon number, displaying limit-cycle oscillations in the long-time limit. (b) Bloch spheres depicting the long-time dynamics (blue) of the collective spins preceding the start of the steady-state limit cycle (thick red line). Insets: (i) Magnified picture of the limit-cycle oscillations; (ii) Time evolution for $\phi=0.4$, corresponding to the NP within adiabatic elimination, but unstable when including the cavity field dynamics.}
\end{figure}

The most striking difference with the ground-state phase diagram is the emergence of an unstable region inside the NP. To understand this, we look at the spectrum of the dynamical matrix in Eq.~\eqref{eq:er0012}, which in the NP ($S_{x,\pm1}=0$) reads
\begin{equation}
\label{eq:er0014}
\eta_{\pm,\pm}= \pm i\sqrt{\omega_{0}(\omega_{0}+\xi) \pm \omega_{0}\sqrt{\chi_{+}\chi_{-}}},
\end{equation}    
where $\xi=\xi_\pm=\Omega \Delta$ and  $\chi_{\pm}=\Omega[\Delta \cos (2\phi) \mp \frac{\kappa}{2} \sin(2\phi)]$, with $\Omega=\lambda^{2}/(\Delta^{2} + \frac{\kappa^2}{4})$.
From this expression, we see that when the couplings $\chi_{\pm}$ acquire opposite sign,
$\tan^2{(2 \phi)} > 4 \Delta^{2}/\kappa^2$, the imaginary parts (frequencies) coalesce, while the real parts (decay rates) become finite, resulting in the emergence of decay and anti-damping, see Fig.~\ref{Spectrum}(a). This phenomenon is known as level attraction and can only arise in non-Hermitian matrices, such as the dynamical matrix in Eq.~\eqref{eq:er0012} \cite{Moiseyev}. This is signaled by the presence of exceptional points, where the eigenvalues are degenerate and the eigenvectors coalesce. In our case, this corresponds to $\tan^2{(2 \phi)} = 4 \Delta^{2}/\kappa^2$, where we have $\chi_{\pm}=0$.
The emergent anti-damping is what makes the NP unstable.
This can be observed in Fig.~\ref{Dynamics}, where we show the time-evolution of the cavity field, from the semi-classical equations of motion \cite{SM}. Initializing the system in the NP with small fluctuations in the cavity field, we see how the system does not remain in this phase, but becomes unstable and at long times features a limit cycle
\footnote{
We have used parameters different from those in the recent experiment \cite{EsslingerSpin} for the phase diagram in Fig.~\ref{PDOrdPar}(b), so that clean self-sustained oscillations are established within the integration times available through numerical integration.
}.

We trace the origin of the unstable behavior to the cavity field mediating non-reciprocal interactions. 
The form of the couplings follows from the equations of motion of the cavity field in the NP, yielding $\chi_{\pm}=-\frac{i}{2}\lambda_{\pm1}^{*}\lambda_{\mp1}\chi_{R}+\frac{i}{2}\lambda_{\pm1} \lambda_{\mp1}^{*}\chi_{R}^{*}$, with $\chi_{R}=(-i\Delta+\kappa/2)^{-1}$ the cavity response function in the bad-cavity limit. 
Each term represents an amplitude for a photon scattering process from one spin to another, with $\chi_{\pm}$ being the total scattering amplitude for each pathway.
For $\kappa=0$, the total scattering amplitudes are symmetric under the exchange of spins ($\chi_{+}=\chi_{-}$).
Conversely, for finite $\kappa$, the phase shift induced by the cavity response results in the interference between the scattering amplitudes being different for each pathway, leading to a non-reciprocal coupling. 
Thus, we can conclude that emergence of the dynamical instability is a consequence of the dissipative nature of the cavity field.
Note that, nevertheless, the phase difference between $\lambda_{\pm1}$ is a crucial ingredient, as for $\phi=0$ both pathways are equivalent, independently of the value of $\kappa$.
This constitutes one of the major findings of this work, which should be contrasted with the impact of photon loss in the standard Dicke model. There, it leads to a shift of the critical point \cite{Bhaseen1,DomokosPRA2011,Oztop2012,Bhaseen2,Dimer2007} and a change in the critical exponent \cite{DomokosPRA2011}, but the ground-state and steady-state phase diagrams are qualitatively similar.

For $\lambda >\lambda^{\textrm{SS}}_{\textrm{cr}}=\sqrt{\frac{(-\omega_{0})\left( \Delta^{2} + \kappa^2/4 \right)}{\Delta\pm\sqrt{\Delta^{2}\cos^{2}{(2\phi)} - (\kappa^2/4)\sin^{2}{(2\phi)} }}}$, the system becomes unstable favoring two different steady-state superradiant phases, which we denote as SDW I  and SDW II. These are different from the DW and SW phases in Fig.~\ref{PDOrdPar}(a) as the effects of dissipation in the steady-state equations leads to $|S_{x,1}| \neq |S_{x,-1}|$, resulting in simultaneous presence of density- and spin-wave order. The SDW I phase is a reminiscent of the DW phase with two of solutions corresponding to the spins being almost aligned and the SDW II phase is instead reminiscent of the SW phase, with a pair of solutions associated with the spins being almost anti-aligned.
Finally, we identify a fifth phase at large coupling $\lambda$, where both SDW I and II are steady states of the system. This is analogous to the top middle part in Fig.~\ref{PDOrdPar}(a) where both the DW and SW phases are local minima of the mean-field energy.

\textit{Beyond adiabatic elimination.---} Finally, we go beyond adiabatic elimination and include cavity field fluctuations. The steady-state equations for the cavity field $\alpha$ and spins $S_{x,\pm1}$ remain unchanged.
In the linear stability analysis we now have to include the dynamics of the cavity-field fluctuations $\delta \hat{a}$ and $\delta \hat{a}^{\dagger}$,
leading to a dynamical matrix that is a 6$\times$6 matrix
which does not allow for an analytical expression for the eigenvalues.
By solving these equations numerically, we find that the resulting phase diagram is qualitatively similar to the one presented in Fig.~\ref{PDOrdPar}(b), with the important exception of the NP being unstable for all $\phi \neq 0,\pm\frac{\pi}{2}$ \cite{SM}.
This is due to a pair of complex conjugate eigenvalues $\varepsilon^{\pm}_{+}$ with finite real positive part.
In Fig.~\ref{Spectrum}(b), we show the real part of these eigenvalues.
These are finite for all $\phi \neq 0,\pm\frac{\pi}{2}$ and they reduce to expression (\ref{eq:er0014}) we obtained above in the limit $(|\Delta|,\kappa) \gg (\omega_{0},\lambda)$. As a consequence the system is driven into limit cycles all throughout the region associated with the solutions $S_{x,\pm1}=0$. This is shown in the inset of Fig.~\ref{Dynamics}, where the time-evolution is considered at a point where adiabatic elimination predicts the NP to be stable, i.e.~$\lambda < \lambda_{\textrm{cr}}^{\textrm{SS}}$ and $\tan^2{(2\phi)} < 4 \Delta^{2}/\kappa^2$. We observe how the system initially remains in the NP, but at longer times, the system dynamics features limit-cycle behavior.

We investigate this further by calculating the eigenvalues perturbatively, exploiting that this phenomenon is also present at small $\lambda$.
In Fourier space, the linearized equations of motion for the cavity fluctuations are 
\begin{equation}
\label{eq:er0015a}\chi_{R}^{-1}(\omega)\delta\hat{a}(\omega)
=\frac{-i\lambda }{2 \sqrt{N}} \sum_{\sigma}
\left[ \delta \hat{J}^{+}_{\sigma}(\omega)  + \delta \hat{J}^{-}_{\sigma}(\omega) \right]
e^{i\sigma\phi} -\sqrt{\kappa}\hat{a}_{\textrm{in}}
\end{equation}
with $\left[\delta \hat{a}(\omega)\right]^{\dagger}=\delta\hat{a}(-\omega)^{\dagger}$, $\chi_{R}(\omega)=[-i(\Delta+\omega)+\kappa/2]^{-1}$ the cavity response function at finite frequency, $\hat{J}^{\pm}=\hat{J}_{x} \pm i\hat{J}_{y}$ the spin raising and lowering operators and $\hat{a}_{\textrm{in}} $ the cavity input noise.
For $\lambda=0$, $\delta\hat{J}^{+}_{\pm1}$ and $\delta\hat{J}^{-}_{\pm1}$ rotate with frequencies $\omega_{0}$ and $-\omega_{0}$, respectively. Using a rotating-wave approximation, we can consider these pairs of modes to be effectively uncoupled and focus only on the dynamics of $\delta \hat{J}^{-}_{\pm1}$.
Substituting Eq.~\eqref{eq:er0015a} into the equation of motion for $\delta \hat{J}^{-}_{\pm1}(\omega)$, we obtain
\begin{equation}
\label{eq:er0016}
\begin{pmatrix} i (\omega_{0}-\omega) + \frac{i}{2}\Sigma (\omega_{0}) & \frac{i}{2}\Lambda_{+}(\omega_{0}) \\ \frac{i}{2}\Lambda_{-}(\omega_{0}) & i (\omega_{0}-\omega) + \frac{i}{2}\Sigma (\omega_{0}) \end{pmatrix} \begin{pmatrix}   \delta \hat{J}^{-}_{1}  \\ \delta \hat{J}^{-}_{-1}  \end{pmatrix} = \hat{\vec{\Gamma}}_{\textrm{in}}
\end{equation}  
with $\Sigma(\omega)=\frac{\lambda^2}{2}[-i\chi_{R}(\omega)+i\chi_{R}^{*}(-\omega)]$ the self-energy and $\Lambda_{\pm}(\omega)=\frac{\lambda^2}{2}[-i\chi_{R}(\omega)e^{\mp 2i \phi}+i\chi_{R}^{*}(-\omega)e^{\pm 2i \phi}]$ the non-reciprocal coupling,
where in the sprit of Fermi's Golden Rule we have evaluated the energy-dependent self-energy and coupling at the unperturbed frequency of the mode $\omega_0$ \cite{Marquardt2007}. We have incorporated all noise terms in $\hat{\vec{\Gamma}}_{\textrm{in}}$. 
The spectrum follows from the determinant of the dynamical matrix \eqref{eq:er0016} as
\begin{equation}
\label{eq:er0017}
\varepsilon^{-}_\pm= -i\omega_{0} - \frac{i}{2}\Sigma(\omega_{0}) \pm \frac{i}{2} \sqrt{\Lambda_{+}(\omega_{0})\Lambda_{-}(\omega_{0})}.
\end{equation}
The self-energy $\Sigma(\omega_{0})$ only provides a frequency shift and a finite damping rate, given that for $\Delta<0$, $\textrm{Im}[\Sigma(\omega_{0})]<0$.
On the contrary, the couplings $\Lambda_{\pm}(\omega_{0})$ always yield an anti-damping contribution, which cannot be compensated by the self-energy damping, as they emerge in a $\pm$ pair and due to $\textrm{Im}[\Lambda_{+}(\omega_{0})\Lambda_{-}(\omega_{0})] \neq 0$ for all $\phi \neq 0,\pm\frac{\pi}{2}$ and $\kappa \neq 0$. Thus, the finite-frequency response of the cavity fluctuations is responsible for the emergence of anti-damping and for the NP becoming unstable to self-sustained oscillations.
As expected, in the limit $\kappa \rightarrow 0$,  we obtain $\textrm{Im}[\Lambda_{+}(\omega_{0})\Lambda_{-}(\omega_{0})] \rightarrow 0$ and $\textrm{Re}[\Lambda_{+}(\omega_{0})\Lambda_{-}(\omega_{0})] > 0$, thus restoring the stability of the NP and confirming again the dissipative nature of the instability. Interestingly, in this limit, the interaction still remains non-reciprocal $\Lambda_{+}(\omega_{0}) \neq \Lambda_{-}(\omega_{0})$, meaning that outside the bad-cavity regime the presence of non-reciprocity does not imply unstable behavior. A second pair of eigenvalues $\varepsilon^{+}_\pm$ is obtained if one considers the dynamics of $\delta \hat{J}^{+}_{\pm1}$ instead, which together with \eqref{eq:er0017} provides an approximate form for the eigenvalues $\varepsilon^{\pm}_{+}$ shown in Fig.~\ref{Spectrum}(b) in the limit of small $\lambda$.
The form of \eqref{eq:er0017} also explains why in the bad-cavity limit the instability is confined to a finite region. More specifically, the bad-cavity limit is equivalent to $\omega \rightarrow 0$, corresponding to the zero-frequency response of the cavity fluctuations being the only component playing a role in the dynamics. This leads to $\Sigma(\omega) \rightarrow \xi$ and $\Lambda_{+}(\omega)\Lambda_{-}(\omega) \rightarrow \chi_{+} \chi_{-}$, i.e.~the instability occurs if $\chi_{+}\chi_{-}<0$, in agreement with our previous result. 

\textit{Outlook.---} Our work opens exciting avenues for future investigations. First, finding an exact solution similar to \cite{Emary} or efficient numerics \cite{Shammah2018} would allow one to explore the instability beyond the semi-classical approximation employed here. Second, non-reciprocity has recently been investigated with several platforms \cite{Jalas2013, Lodahl2017, Bernier2017, Verhagen2017} that have been specifically engineered. Here, it emerges naturally as a consequence of the dissipative nature of the cavity field, offering a testbed for non-reciprocal phenomena in a highly-controlled environment. In particular, interesting directions include the impact of non-reciprocity on higher-order photon correlations \cite{Harder2018} or on synchronization behavior \cite{Xu2014}. Third, the effects of interatomic interactions can be investigated with an additional optical lattice and lead to complex many-body behavior \cite{Mazzucchi2016}. Finally, following Refs.~\cite{EsslingerSpin,Kroeze2018}, we expect the emergence of the unstable regime and the steady-state phase diagram in Fig.~\ref{PDOrdPar} to be experimentally observable,  by means of photon counting and heterodyne detection.

\begin{acknowledgments}
We thank Matteo Brunelli for careful reading of the manuscript, Tobias Donner for stimulating discussions that motivated this work, and Christopher Parmee for useful comments on the numerical calculations. E.I.R.C.~acknowledges the Winton Programme for the Physics of Sustainability and the UK Engineering and Physical Sciences Research Council (EPSRC) under Grant No.~EP/N509620/1. A.N.~acknowledges a University Research Fellowship from the Royal Society and the Winton Programme for the Physics of Sustainability.
\end{acknowledgments}

\textit{Note added.---} Recently, a preprint \cite{Dogra2019} has reported the observation of the dissipation-induced instability we discuss.

\bibliographystyle{apsrev4-1}
\bibliography{library}

\newpage
\clearpage
\appendix
	
\setcounter{figure}{0}
\makeatletter 
\renewcommand{\thefigure}{S\arabic{figure}}
	
\newcounter{defcounter}
\setcounter{defcounter}{0}
	
\newenvironment{myequation}
{%
\addtocounter{equation}{-1}
\refstepcounter{defcounter}
\renewcommand\theequation{S\thedefcounter}
\align
}
{%
\endalign
}
\setcounter{page}{1}
	
\begin{widetext}
\begin{center}
{\fontsize{12}{12}\selectfont
\textbf{Supplemental Material for ``\papertitle''\\[5mm]}}
{\normalsize \authornames\\[1mm]}
{\fontsize{9}{9}\selectfont  
\textit{\tcm}}
\end{center}
\normalsize
\end{widetext}

\section{Mean-field ground-state phase diagram}

In this section we describe the main steps taken to obtain the form of the ground-state phase diagram presented in the main text. This was first discussed in Ref.~\cite{EsslingerSpin}.
We start by introducing the Holstein-Primakoff transformation $\hat{J}_{z,\pm 1}=\hat{b}_{\pm}^{\dagger}\hat{b}_{\pm}-\frac{N}{2}$ and $\hat{J}_{x,\pm 1}=\frac{1}{2} \left(\hat{b}_{\pm}^{\dagger}\sqrt{N-\hat{b}_{\pm}^{\dagger}\hat{b}_{\pm}} + \sqrt{N-\hat{b}_{\pm}^{\dagger}\hat{b}_{\pm}}\: \hat{b}_{\pm} \right)$ and inserting it in the Hamiltonian (\ref{eq:er0002}) defined in the main text, which leads to (up to a constant shift)
\begin{align}
\label{eq:erSI0001}
\hat{H}=&-\Delta \hat{a}^{\dagger} \hat{a} + \sum_{\sigma = \pm} \Big[ \omega_0 \hat{b}_{\sigma}^{\dagger} \hat{b}_{\sigma}+\frac{\lambda}{2}\left(\hat{a} e^{-i\sigma\phi} + \hat{a}^{\dagger} e^{i\sigma\phi}\right)  \nonumber \\ 
&\times\left(\hat{b}_{\sigma}^{\dagger}\sqrt{N-\hat{b}_{\sigma}^{\dagger}\hat{b}_{\sigma}}  + \sqrt{N-\hat{b}_{\sigma}^{\dagger}\hat{b}_{\sigma}} \: \hat{b}_{\sigma}\right) \Big].
\end{align}
We obtain the ground-state phase diagram of the Hamiltonian \eqref{eq:erSI0001} by studying the mean-field energy 
\begin{align}
\label{eq:erSI0001b}
E_{\textrm{MF}}= & -\Delta |\alpha|^2 +\sum_{\sigma=\pm} \Big[ \omega_0 |\beta_{\sigma}|^2   \nonumber \\
&+ \lambda(\alpha e^{-i\sigma\phi} + \alpha^{*}e^{i\sigma\phi})\textrm{Re}[\beta_{\sigma}]\sqrt{1-\frac{|\beta_{\sigma}^{2}|}{N}}\Big],
\end{align}
where $\alpha=\langle \hat{a} \rangle$ and $\beta_{\pm} = \langle \hat{b}_{\pm} \rangle$. Minimizing the energy with respect to both fields yields $\beta_{\pm}=\textrm{Re}[\beta_{\pm}]$ and $\alpha=\frac{\lambda}{\Delta} \sum_{\sigma} e^{i\sigma\phi} \beta_{\sigma}\sqrt{1-\frac{\beta_{\sigma}^{2}}{N}}$. We expand the square-root terms to first order, since close to the phase transition $\frac{\beta_{\pm}}{N} \ll 1$, and introduce new order parameters $\beta_{d,s}=\frac{1}{\sqrt{2}}(\beta_{+} \pm \beta_{-})$, which signal the presence of density- and spin-wave order, respecitvely. This leads to 
\begin{align}
\label{eq:erSI0002}
E_{\textrm{MF}} \simeq & \left[ \omega_{0} +\frac{\lambda^2}{\Delta}(1+\cos{(2\phi)}) \right] \beta_{d}^2 - \frac{\lambda^{2}}{2N\Delta}(1+\cos{(2\phi)})\beta_{d}^{4} \nonumber \\
+&\left[ \omega_{0} +\frac{\lambda^2}{\Delta}(1-\cos{(2\phi)}) \right] \beta_{s}^2 - \frac{\lambda^{2}}{2N\Delta}(1-\cos{(2\phi)})\beta_{s}^{4}.
\end{align}
From the global minima of \eqref{eq:erSI0002}, we distinguish three different phases as function of $\lambda$ and $\phi$, see Fig.~\ref{PDOrdPar}(a) in the main text. First, for $\lambda<(\lambda_{\textrm{cr}}^{d},\lambda_{\textrm{cr}}^{s})$, with $\lambda_{\textrm{cr}}^{d}=\sqrt{\frac{(-\Delta)\omega_{0}}{1+\cos{(2\phi)}}}$ and $\lambda_{\textrm{cr}}^{s}=\sqrt{\frac{(-\Delta)\omega_{0}}{1-\cos{(2\phi)}}}$, both order parameters vanish $\beta_{d,s}=0$, corresponding to the NP, where both collective spins point down in the $z$ direction, meaning that all atoms remain in the BEC state.
Second, for $\lambda>\lambda_{\textrm{cr}}^{d}$ and $0 \leq \phi \leq \frac{\pi}{4}$, we obtain $\beta_{d}=\pm\sqrt{N(1-(\lambda_{\textrm{cr}}^{d}/\lambda)^{2})}$, while $\beta_{s}=0$. This is associated with both collective spins aligning in the $x$-$z$ plane, meaning that both atomic species self-organize in the same pattern, thus realizing a density-wave state (DW). Lastly, for $\lambda>\lambda_{\textrm{cr}}^{s}$ and $\frac{\pi}{4} \leq \phi \leq \frac{\pi}{2}$, we obtain $\beta_{s}=\pm\sqrt{N(1-(\lambda_{\textrm{cr}}^{s}/\lambda)^{2})}$ and $\beta_{d}=0$, where now the $x$-components of the spins point in opposite directions. This can be understood as the atomic species self-organizing in the opposite checkerboard patterns, resulting in the emergence of a spin-wave state (SW). 

\section{Steady-state equations}

The semi-classical equations of motion for the system in the presence of cavity loss are
\begin{align}
\partial_{t} \alpha &= \left(i \Delta - \frac{\kappa}{2} \right) \alpha -i\frac{\lambda}{\sqrt{N}} \left(S_{x, 1} e^{i\phi} + S_{x,-1} e^{-i\phi} \right) \nonumber \\
\partial_{t} S_{x, \pm 1} &= - \omega_0 S_{y, \pm 1} \nonumber \\
\partial_{t} S_{y, \pm 1} &= \omega_0 S_{x, \pm 1} - \frac{\lambda}{\sqrt{N}}
\left(\alpha e^{\mp i\phi} + \alpha^{*} e^{\pm i\phi} \right)S_{z, \pm 1} \nonumber \\
\label{eq:erSI0003}
\partial_{t} S_{z, \pm 1} &=  \frac{\lambda}{\sqrt{N}} \left(\alpha e^{\mp i\phi} + \alpha^{*} e^{\pm i\phi} \right)S_{y, \pm 1}.
\end{align}

Setting $\alpha \approx  \frac{\lambda}{\sqrt{N}}\frac{S_{x,1} e^{i\phi}+S_{x,-1} e^{-i\phi}}{\Delta + i\frac{\kappa}{2}}$ and $\partial_{t} S_{\sigma, \pm 1}=0$, leads to a set of algebraic equations that define the steady-state solutions for the spins
\begin{align}
\label{eq:erSI0004}
&S_{x,\pm 1} + \frac{2 \lambda^{2}}{N(\Delta^{2} + \frac{\kappa^2}{4})}
\left\{ \Delta S_{x,\pm 1} + \left[ \Delta \cos{(2\phi)} \right. \right. \nonumber \\
& \left. \left. \mp \frac{\kappa}{2} \sin{(2\phi)}\right] S_{x, \mp 1} \right\} \sqrt{\frac{N^2}{4} - S_{x,\pm 1}^{2} }=0 ,
\end{align}
where we used $S_{y,\pm 1}=0$ and $S_{z, \pm 1} = - \sqrt{\frac{N^2}{4} - S_{x,\pm 1}^2}$.

The linearized equations of motion in the main text follow from inserting the expansion $\hat{J}_{\sigma,\pm 1}(t) \simeq S_{\sigma,\pm 1} + \delta \hat{J}_{\sigma, \pm 1}(t)$ in the equations of motion \eqref{eq:erSI0003} and keeping only the terms which are linear in $\delta \hat{J}_{\sigma, \pm 1}(t)$, where the $z$-component is fixed by spin conservation $S_{z,\pm1} \delta \hat{J}_{z,\pm1}=-S_{x,\pm1} \delta \hat{J}_{x,\pm1}$. We obtain the following expression for the self-energy and the coupling
\begin{align}
\label{eq:erSI0005a}
\xi_{\pm}&=\frac{2 \lambda^{2} S_{z, \pm 1}}{N(\Delta^{2} + \frac{\kappa^2}{4})} \left\{\Delta \left( \frac{S^{2}_{x, \pm 1}}{S^{2}_{z, \pm 1}}-1\right) \right.\nonumber \\
&\left.+\left[\Delta \cos{(2 \phi)}) \mp \frac{\kappa}{2} \sin{(2 \phi)}\right] \frac{S_{x,1}S_{x,-1}}{S^{2}_{z, \pm 1}} \right\} \\
\label{eq:erSI0005b} \chi_{\pm}&=-\frac{2 \lambda^{2} S_{z, \pm 1}}{N(\Delta^{2} + \frac{\kappa^2}{4})}
\left[\Delta \cos{(2 \phi)} \mp \frac{\kappa}{2} \sin{(2 \phi)}\right].
\end{align}

\section{Analysis beyond adiabatic elimination}

Outside the bad-cavity regime, to obtain the phase diagram, we need to perform a stability analysis of the solutions of Eq.~\eqref{eq:erSI0004} taking into account cavity fluctuations $\delta \hat{a}$ and $\delta \hat{a}^{\dagger}$.
The linearized equations of motion read\\
\\
\\
\\
\begin{widetext}
\begin{equation}
\label{eq:erSI0006}
\begin{pmatrix} \delta \dot{\hat{a}} \\ \delta \dot{\hat{a}}^{\dagger} \\ \delta \dot{\hat{J}}_{x,1} \\ \delta \dot{\hat{J}}_{y,1} \\ \delta \dot{\hat{J}}_{x,-1} \\ \delta\dot{\hat{J}}_{y,-1} \end{pmatrix}=
\begin{pmatrix}
 i\Delta - \frac{\kappa}{2} & 0 & -i\frac{\lambda}{\sqrt{N}} e^{i\phi} & 0 & -i\frac{\lambda}{\sqrt{N}} e^{-i\phi} & 0 \\
 -i\Delta - \frac{\kappa}{2} & 0 & i\frac{\lambda}{\sqrt{N}} e^{-i\phi} & 0 & i\frac{\lambda}{\sqrt{N}} e^{i\phi} & 0 \\
 0 & 0 & 0 & -\omega_{0} & 0 & 0 \\
-\frac{\lambda}{\sqrt{N}} S_{z,1} e^{-i\phi} & -\frac{\lambda}{\sqrt{N}} S_{z,1} e^{i\phi} & \omega_{0} + \Xi_{+} & 0 & 0 &0 \\
 0 & 0 & 0 & 0 & 0 & -\omega_{0} \\
-\frac{\lambda}{\sqrt{N}} S_{z,-1} e^{i\phi} & -\frac{\lambda}{\sqrt{N}} S_{z,-1} e^{-i\phi} & 0 & 0 & \omega_{0} + \Xi_{-} & 0
\end{pmatrix}
\begin{pmatrix} \delta \hat{a} \\ \delta \hat{a}^{\dagger} \\ \delta \hat{J}_{x,1} \\ \delta \hat{J}_{y,1} \\ \delta \hat{J}_{x,-1} \\ \delta \hat{J}_{y,-1} \end{pmatrix},
\end{equation}
\end{widetext}
where $\Xi_{\pm}$ is related to $\xi_{\pm}$, as defined in Eq.~\eqref{eq:erSI0005a}, through $\Xi_{\pm}=\xi_{\pm}+ \frac{2 \Delta \lambda^{2} S_{z, \pm 1}}{N(\Delta^{2} + \frac{\kappa^2}{4})}$.  The eigenvalues of this dynamical matrix cannot be obtained analytically, independently of the values of the order parameters, and thus need to be computed numerically. The eigenvalues $\varepsilon^{\pm}_{+}$ plotted in the main text are obtained by numerically diagonalizing the dynamical matrix setting $S_{x,\pm1}=0$ and $S_{z,\pm1}=-\frac{N}{2}$. As in depicted in Fig.~\ref{PDOrdPar}(b) in the main text, due to the cavity fluctuation dynamics, the unstable region is enhanced, with the NP remaining stable only for $\phi=0$ and $\phi=\pm\frac{\pi}{2}$.

\end{document}